\documentclass[12pt]{iopart}
\usepackage{graphics,graphicx,subfigure}

\begin{document}

\title{Two-dimensional Wigner crystals of classical Lennard-Jones particles}

\author{Igor Trav\v{e}nec and Ladislav \v{S}amaj}

\address{Institute of Physics, Slovak Academy of Sciences, 
	D\'ubravsk\'a cesta 9, 84511 Bratislava, Slovakia}
\ead{fyzitrav@savba.sk}
\vspace{10pt}
\begin{indented}
\item[]
\end{indented}

\begin{abstract}
The ground-state of two-dimensional (2D) systems of classical particles 
interacting pairwisely by the generalized Lennard-Jones potential is studied.
Taking the surface area per particle $A$ as a free parameter and 
restricting oneself to periodic Bravais lattices with one particle per unit
cell, B\'etermin L [2018 Nonlinearity {\bf 31} 3973] proved that the 
hexagonal, rhombic, square and rectangular structures minimize successively 
the interaction energy per particle as $A$ increases.
We show here that the second-order transitions between the rhombic/square
and square/rectangular phases are of mean-field type.
The aim of this paper is to extend B\'etermin's analysis to periodic 
2D lattices with more than one particle per elementary cell.
Being motivated by previous works dealing with other kinds of models, 
we propose as the ground-state the extensions of the 2D rectangular (1-chain)
lattice, namely the ``zig-zag'' (2-chain), 3-chain, 4-chain, etc. structures 
possessing 2, 3, 4, etc. particles per unit cell, respectively.
By using a recent technique of lattice summation we find for the standard 
Lennard-Jones potential that their ground-state energy per particle 
approaches systematically as the number of particles per unit cell increases
to the one of a phase separated state (the optimal hexagonal lattice).
We analyze analytically the low-density limit $A\to\infty$ and 
the limiting hard-core case of the generalized Lennard-Jones potential.
\end{abstract}


\vspace{2pc}

\submitto{\JPA}

\maketitle

\section{Introduction}
Systems of classical pointlike particles interacting pairwisely via 
Lennard-Jones potential (LJ) \cite{L-J} have been studied intensively.
For two particles at distance $r$, the generalized version of this potential
\begin{equation} \label{n-m}
V(r) = \frac{1}{n-m}\left(\frac{m}{r^{n}}-\frac{n}{r^m}\right), \quad n>m 
\end{equation}
is often referred to as the Mie potential.
Here, $n$ and $m$ are positive integers, the inequality $n>m$ prevents 
the collapse of particle pairs in the ground state.
The first repulsive term is dominant at small distances, the second attractive
term prevails at large distances between the particles. 
Since the particle distance with the minimum energy is given by $r_{\min}=1$ and 
$V(1)=-1$, the length and energy can be taken in arbitrary units.
The potential (\ref{n-m}) is denoted as the $n-m$ LJ potential. 

We will concentrate on the choice $m=6$, i.e. the $n-6$ model. 
This combination is considered very often, see e.g. \cite{Sousa}, 
because $1/r^6$ is the attractive long-range van der Waals potential 
between molecules with electric dipole moments \cite{Milton}.
The original LJ value for the repulsive $n=12$ potential applies 
to argon \cite{Verlet}, but some other rare gases have an effective value 
of $n$ close to $11$.
For $n$ finite, we speak about the soft potential since the particles can 
approach to one another arbitrarily close.
In the $n\to\infty$ limit, we get the hard-core potential of diameter 1
around each particle with an attractive interaction added,
\begin{equation} \label{hard-core}
V(r) = \cases{\infty \qquad r<1, \cr
-\frac{1}{r^m} \quad r\ge 1.}
\end{equation}
Values of $n$ up to $n=36$ were studied in Ref. \cite{Sousa}.
Numerous other constraints for $m$ and $n$ were studied. 
The case $n=2m$ seems to be motivated by the most common $12-6$ potential
rather than by a convincing physical situation.
Nevertheless the $24-12$ and $36-18$ cases mimic, at least locally,
colloidal systems during phase separation \cite{Lodge}.
For systems in higher $d$-dimensions, the choice $n=d+9$ and $m=d+3$ 
was suggested for 3D gases.
Sometimes the $9-3$ potential is used for the solid-fluid interaction model 
\cite{Smith}.
The special case $n=6k-3$, $m=3k-3$ was found to be adequate for solids 
satisfying the spinodal condition \cite{Sun}.
The treatment of dynamics and pattern formation in systems with competing 
interactions or an external force can be found in Ref. \cite{ZMP}.

As has been already mentioned, the potential (\ref{n-m}) is attractive at 
large distances.
Hence, at zero temperature, if we take a finite number of particles and 
gradually enlarge the space at their disposal, starting from some volume 
the particles will occupy only a restricted space region and the rest remains 
empty \cite{Gardner,Blanc}.
In the two-dimensional (2D) space, particles create a phase separated state  
(the ``optimal'' hexagonal lattice); the hexagonal structure is known 
to minimize the energy for majority of purely repulsive pair interactions, 
ranging from the long-range Coulomb potential to hard disks.
This phenomenon does not occur if external fields are applied, e.g. a periodic 
field with maxima at the vertices of the square lattice 
\cite{PMZP,Granz,P-T,PZMP}, random pinning \cite{PZMP} or particles are 
restricted to a space domain by geometry, e.g. Wigner bilayers on parallel 
planes \cite{ST,TS2}. 

Another possibility is to consider periodic structures with a fixed area $A$ 
per particle (i.e., the particle density equals to $1/A$) and to minimize 
the interaction energy per particle with respect to those structures. 
B\'etermin \cite{Bet} restricted himself to the standard 2D Bravais lattices 
with one particle per unit cell and applied this approach to the original 
$12-6$ LJ potential. 
He has found that for small $A$, where the repulsive potential dominates 
at short distances, the hexagonal lattice provides the lowest energy.
But as $A$ rises, the rhombic, square and rectangular lattices become 
successively the energy minimizers.

It is natural to extend the ground-state problem to periodic 2D lattices with 
more than one particle per elementary cell.
Our primary motivation comes from the work \cite{Granz} about very distinct 
2D systems of particles, interacting via the dipole $1/r^3$ potential and 
being in an external square-periodic potential. 
For such systems, the structural transition from the square to hexagonal 
lattices is realized via solitonic structures and the double-periodic 
2-chain zig-zag lattice with two particles per unit cell.
It is shown that this zig-zag phase provides a substantially lower ground-state 
energy than simple Bravais structures for intermediate and large values $A$. 
A further generalization comes from the study of quasi-one-dimensional
systems of repulsive Yukawa particles in a confining potential \cite{Q1D}.
In dependence on the parametric form of the potential, the particles
form 2-, 3-, etc. chain configurations.
We introduce 2D counterparts of these quasi-one-dimensional structures 
with respectively 2, 3, etc. particles per unit cell and find for 
the standard 12-6 LJ potential that their ground-state energy approaches 
systematically to the one of the optimal hexagonal lattice.  

As a computational tool we apply a recent technique of lattice summations 
which provides a representation of the energy per particle as quickly 
converging series of Misra functions \cite{ST,TS2}.
In contrast to computer simulation methods, the technique requires 
ad-hoc lattice candidates for the ground-state.
On the other hand, close to second-order phase transitions the method
allows to derive the Landau expansion of the ground-state energy in 
the order parameter.
The critical point is then given as the nullity equation for an 
expansion coefficient and can be specified with a prescribed accuracy
by exploring a short computer time.  
Critical exponents can be obtained analytically.
In the ground-state, critical exponents are usually of mean-field type, but
there are exceptions from this rule \cite{Antlanger}. 
In the context of B\'etermin's analysis \cite{Bet} we show analytically that 
the second-order transitions between the rhombic/square and square/rectangular 
phases are of mean-field type.
Furthermore, being motivated by previous works \cite{Granz,Q1D}, for 
the standard $12-6$ LJ potential we consider the 2D 
zig-zag (2-chain), 3-chain and 4-chain structures which provide 
a lower ground-state energy than simple Bravais lattices for intermediate
and large values of $A$.
It is suggested that by increasing the number of particles in elementary
cell the energy goes to the one of the optimal hexagonal lattice. 
Our technique of lattice summation is also very appropriate to study specific
limiting cases where tiny energy differences among different structures 
are not accessible by simulation methods.   
In particular, for the zig-zag phase we analyze the $n-6$ potential up to 
the hard-disk limit $n\to\infty$ or the small particle-density limit 
$A\to\infty$.

The paper is organized as follows.
In section \ref{Sec2} we summarize basic facts about Bravais lattices, 
the double-periodic (2-chain) zig-zag, 3-chain and 4-chain structures 
and write down Misra series representations of their energies per particle.
The phase diagram for the standard $12-6$ LJ potential is constructed 
in section \ref{Sec3}.
The general $n-6$ case of the LJ potential with $n$ going to large values, 
up to the limit of hard disks is the subject of section \ref{Sec4}.
A short recapitulation is presented in section \ref{Sec5}.

\section{Periodic structures and their energy} \label{Sec2}
To calculate the energy per particle for periodic lattice structures, we use 
a recent method applied already to long-range Coulomb \cite{ST,TS} and 
Yukawa systems \cite{TS2}; for those who would like to reproduce our present 
formulas, the cited papers describe the method in detail.
The method involves the application of the gamma identity to each interaction
term; note that each pair of particles shares the corresponding interaction
energy which therefore must be taken with factor $1/2$ when calculating
the energy per particle. 
Subsequently, the lattice sum is expressed in terms of an integral 
over certain products of the Jacobi theta functions with zero argument 
\cite{Gradshteyn}. 
Repeating twice the Poisson summation formula leads to the representation
of the lattice sum as an infinite series of (generalized) Misra functions
\cite{Misra}. 
The series is converging very quickly; for instance, the truncation of 
the series at the 1st, 2nd, 3rd and 4th term reproduces the Coulomb Madelung 
constant for the hexagonal lattice up to $2,5,10,17$ decimal digits, 
respectively.
The computation of one energy value, e.g. by using MATHEMATICA, requires around
one second of the CPU time on the standard PC.
It turns out that the Misra series representation is technically simpler to
derive for our potential (\ref{n-m}) if $n$ and $m$ are even integers, 
and we shall restrict ourselves to this case.

\begin{figure}[]
\centering
\subfigure[]{\label{f1a}
\includegraphics[clip,width=.42\linewidth]{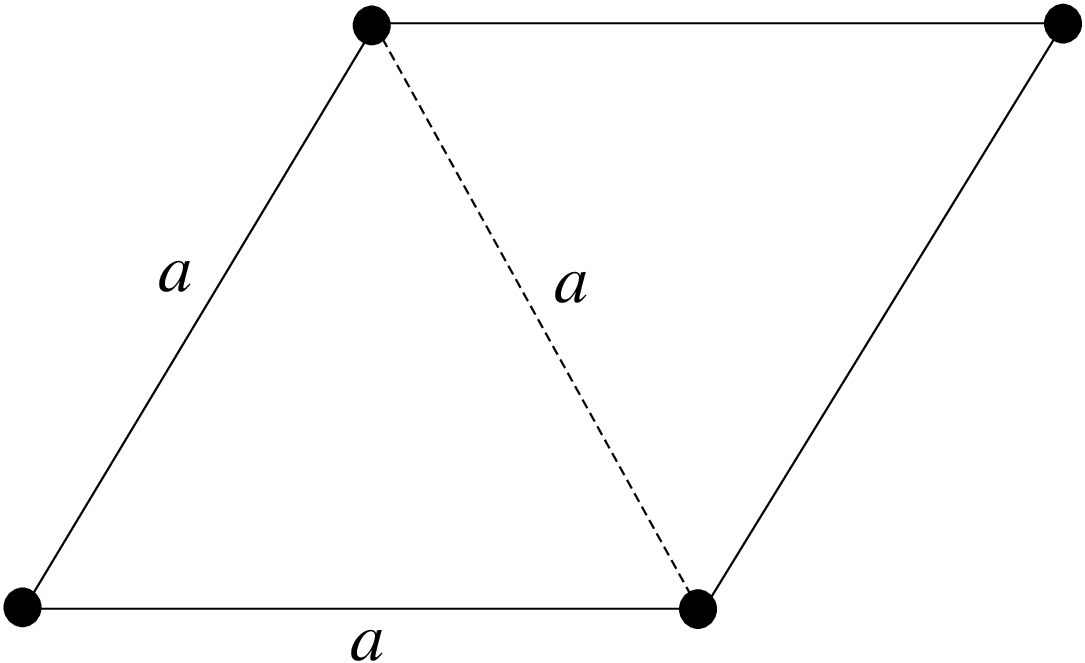}} \quad
\centering
\subfigure[]{\label{f1b}
\includegraphics[clip,width=.34\linewidth]{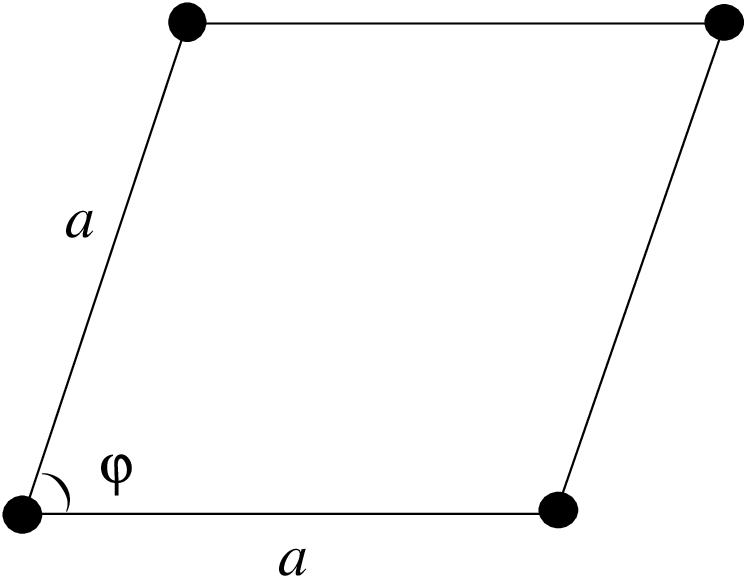}}\\
\centering
\subfigure[]{\label{f1c}
\includegraphics[clip,width=.29\linewidth]{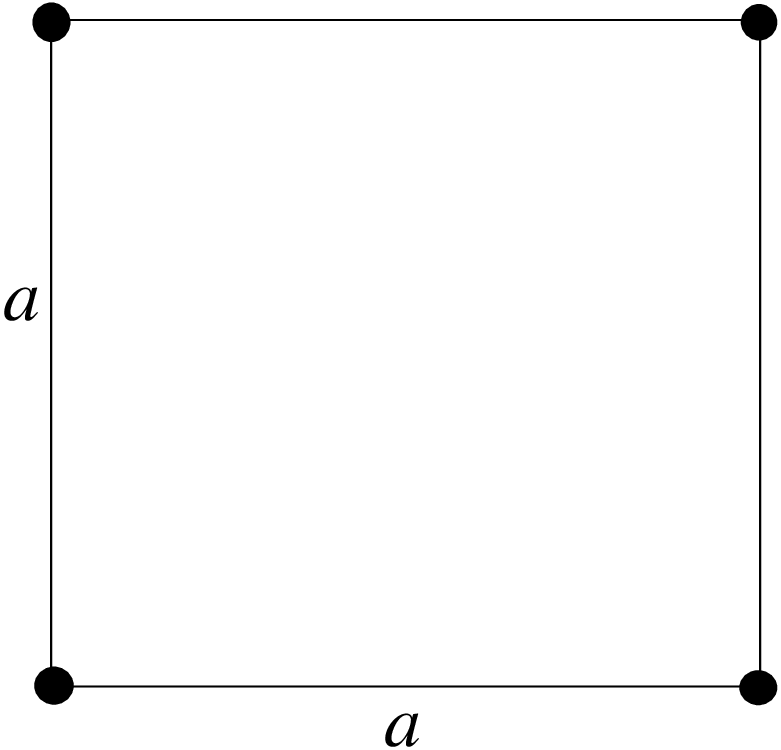}} \quad
\centering
\subfigure[]{\label{f1d}
\includegraphics[clip,width=.4\linewidth]{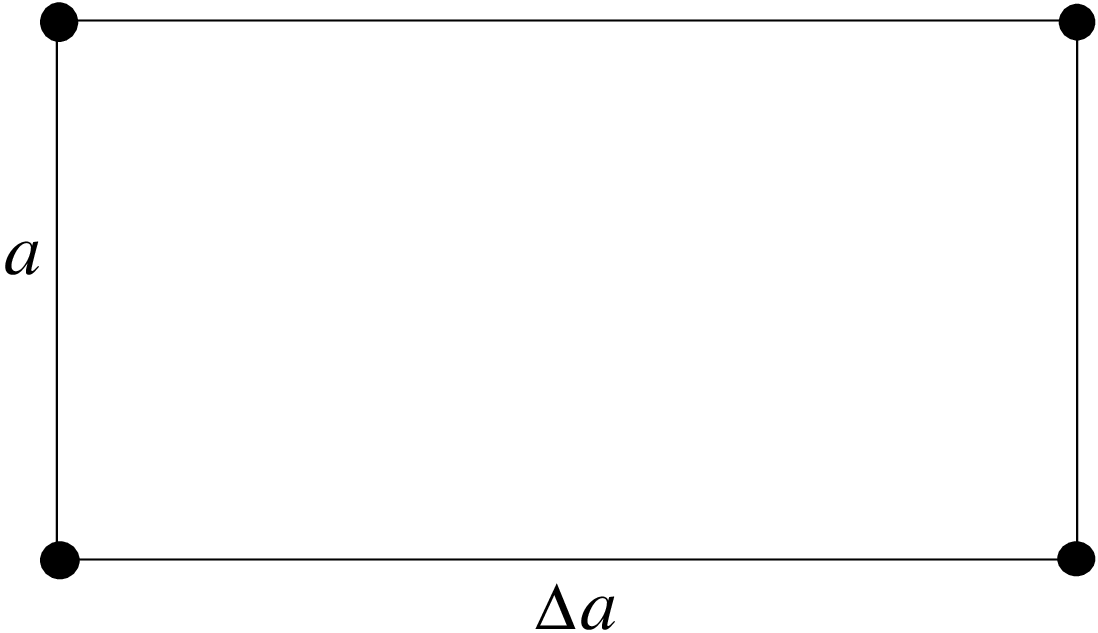}}
\caption {Hexagonal (a), rhombic (b), square (c) and rectangular (d) lattice.}
\label{f1}
\end{figure}

We start with four Bravais lattices with one particle per elementary cell
(see Fig. \ref{f1}) which were treated in Ref. \cite{Bet}.

In the case of the rhombic lattice, the size of each side is equal to $a$, 
the smaller angle is $\varphi$ and the area $A=a^2 \sin{\varphi}$.
We introduce the parameter $\delta=\tan(\varphi/2)$ whose special values 
correspond to two rigid Bravais lattices. 
The choice $\varphi=\pi/3$ (i.e. $\delta=1/\sqrt{3}$), or equivalently
$\varphi=2\pi/3$ (i.e. $\delta=\sqrt{3}$), corresponds to the hexagonal 
lattice, sometimes referred to as the equilateral triangular 
lattice; note that adjacent triangles form a rhomb with the area 
$A=a^2 \sqrt{3}/2$.
For $\varphi=\pi/2$ (i.e. $\delta=1$) one has the square lattice. 
The energy per particle of a rhombic structure with the $n-m$ potential  
(\ref{n-m}) is derived in the form
\begin{equation} \label{enmrh}
E_{n,m}^{\rm rh}(\delta,A) = \frac{1}{n-m}\left[ m \frac{v_n^{\rm rh}(\delta)}{
A^{n/2}}-n \frac{v_m^{\rm rh}(\delta)}{A^{m/2}}\right],
\end{equation}
the Misra series representation of ${v_n^{\rm rh}(\delta)}$ is given (for even 
values of $n$ or $m$) in equation (\ref{vndelta}) of Appendix.
Note the obvious symmetry $\delta\to 1/\delta$ of the rhombic energy which
corresponds to the equivalence of choosing either of the angles $\varphi$ 
or $\pi-\varphi$.

Another structure of interest is the rectangular lattice with aspect ratio 
$\Delta$ and $A=a^2\Delta$.
The energy per particle of this lattice with the $n-m$ potential  
(\ref{n-m}) reads as
\begin{equation} \label{enmrec}
E_{n,m}^{\rm rec}(\Delta,A) = \frac{1}{n-m}\left[ m \frac{v_n^{\rm rec}(\Delta)}{
A^{n/2}}-n \frac{v_m^{\rm rec}(\Delta)}{A^{m/2}}\right] .
\end{equation}
The Misra series representation of ${v_n^{\rm rec}(\Delta)}$ is given
in equation (\ref{vnDelta}) of Appendix. 
It stands to reason that this formula gives the same energy as (\ref{enmrh}) 
for the square lattice, i.e. for $\Delta=\delta=1$.
Note the obvious symmetry $\Delta\to 1/\Delta$ for the rectangular energy.

\begin{figure}[]
\centering
\begin{subfigure}{ \label{fig2a}
\includegraphics[clip,width=.60\linewidth]{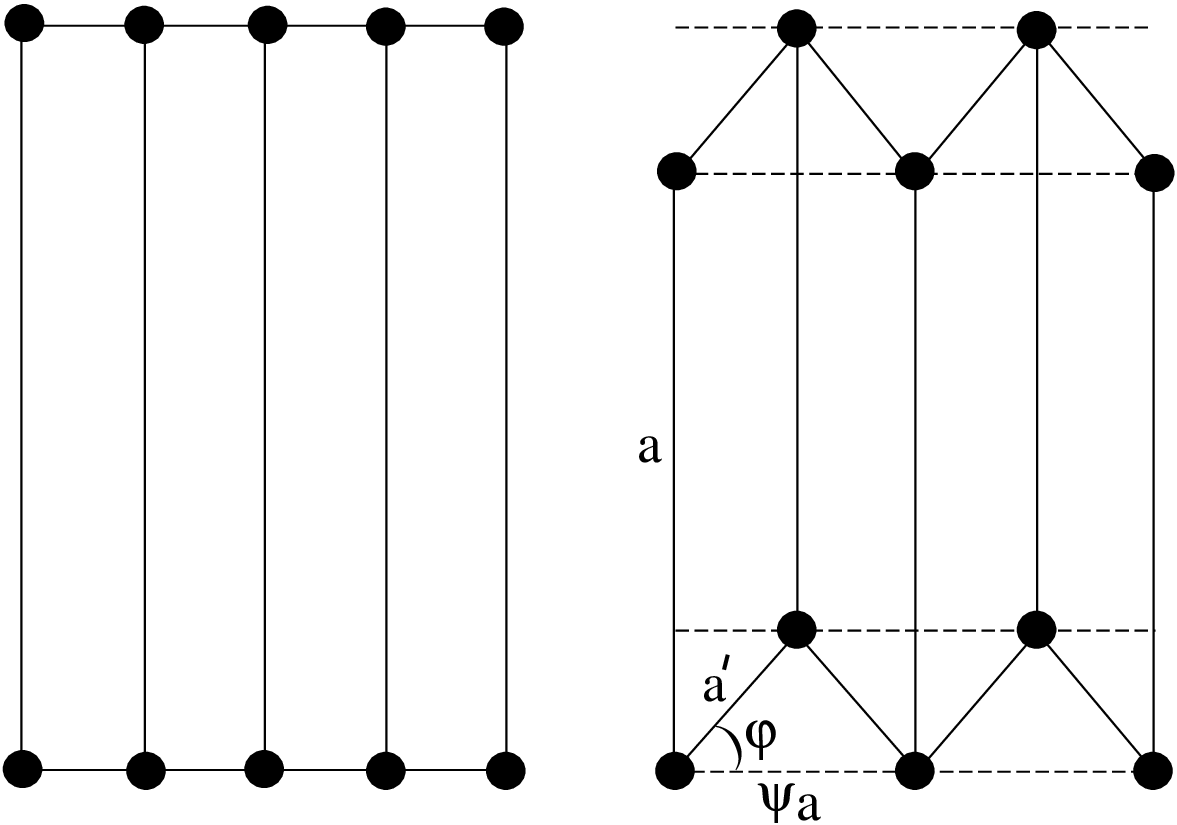}} \qquad
\end{subfigure}
\begin{minipage}{5cm}
\vspace{-6.4cm}
\begin{subfigure}{ \label{fig2b}
\includegraphics[clip,width=.56\linewidth]{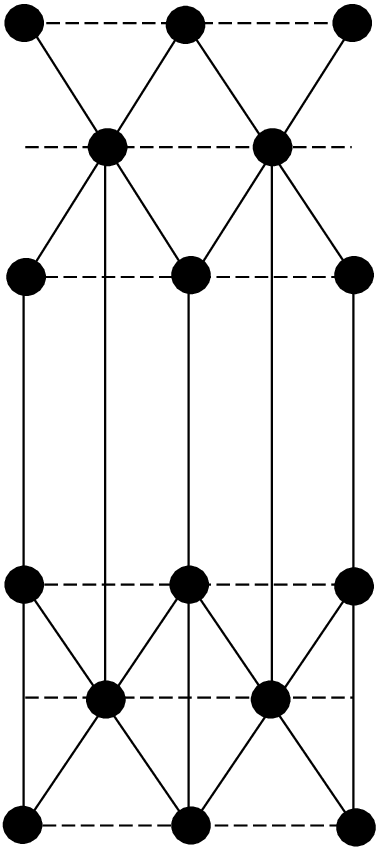}}
\end{subfigure}
\end{minipage}
\caption {The successive extension of the Bravais rectangular (1-chain) 
lattice in the left panel to the zig-zag (2-chain) structure in 
the central panel and the 3-chain lattice in the right panel.
In each case, the elementary cell is given by the rectangle defined by
two vertical solid lines whose ends are connected by two horizontal 
dashed lines.} 
\label{f2}
\end{figure}

Besides simple Bravais lattices there exist other periodic structures 
with more particles per unit cell which can be potential energy minimizers.
Here, we consider the successive more-chain extensions of the Bravais 
rectangular (1-chain) lattice pictured in the left panel of Fig. \ref{f2}. 

The zig-zag (2-chain) structure, adopted from Ref. \cite{Granz}, is presented 
with the notation of lattice parameters in the central panel of Fig. \ref{f2}. 
The elementary cell, given by the rectangle defined by two vertical solid 
lines whose ends are connected by two horizontal dashed lines, contains 
two particles.  
Due to the lattice symmetry, all particles are equivalent in the sense
that they see the same lattice arrangements of other particles.
Since the density of particles is defined as $1/A$, it holds that
\begin{equation} \label{a}
2 A = \psi a^2 .
\end{equation} 
The length $a'$ is given by the relation
\begin{equation} \label{aprime}
\cos\varphi = \frac{\psi a}{2 a'} .
\end{equation}
The special case of the square lattice corresponds to $\varphi=0$, $a=a'$ 
and $\psi=2$.
The hexagonal lattice is identified with the following parameters 
$\varphi=\pi/6$, $a=a'$ and $\psi=2\cos{\varphi}=\sqrt{3}$.
Another possibility for the hexagonal lattice is the parameter choice 
$\varphi=\pi/3$, $a'=\psi a$ and $\psi=1/\sqrt{3}$ which we shall
consider in what follows.
The energy per particle of the zig-zag structure reads as
\begin{eqnarray} \label{ezigzag}
E^{\rm zz}_{n,m}(\varphi,\psi,A)  =  E_{n,m}^{\rm rec}(\psi,2A)
+ E_{n,m}^{\rm shift}\left(\psi,2A,\frac{1}{2},\frac{\psi}{2}\tan{\varphi}\right),
\end{eqnarray}
where the term $E_{n,m}^{\rm shift}(\psi,2A,c_1,c_2)$ sums the energy 
contributions over an infinite rectangular lattice of sides $a$ and $\psi a$, 
the sum being made with respect to a reference point shifted by the vector 
$(c_1\psi a,c_2 a)$ from a lattice point, see equations (\ref{v2}) 
and (\ref{s2}) of Appendix.
For a given value of $A$, the energy (\ref{ezigzag}) is minimized with
respect to the free lattice variables $\varphi$ and $\psi$, 
the lengths $a$ and $a'$ are fixed by the relations 
(\ref{a}) and (\ref{aprime}). 

The 3-chain structure, adopted from Ref. \cite{Q1D}, is presented 
in the right panel of Fig. \ref{f2}. 
The elementary cell, as before given by the central rectangle defined by 
two vertical solid lines whose ends are connected by two horizontal dashed 
lines, contains three particles.  
There are two non-equivalent sets of particles: the first set is given
by two particles inside the unit cell while the second set consists
in four particles forming the ends of the elementary rectangle 
(each weighted by the factor 1/4).
Due to the reflection lattice symmetry, the number of lattice parameters
is the same as in the zig-zag structure and they are defined analogously.
The zig-zag relation (\ref{a}) is modified to
\begin{equation} \label{aa}
3 A = \psi a^2 ,
\end{equation} 
while the relation (\ref{aprime}) remains unchanged.
The energy of the 3-chain structure is obtained in the form
\begin{eqnarray} \label{e3}
E^{\rm 3ch}_{n,m}(\varphi,\psi,A) & = & E_{n,m}^{\rm rec}(\psi,3A)
+ \frac{4}{3}\ E_{n,m}^{\rm shift}
\left(\psi,3A,\frac{1}{2},\frac{\psi}{2}\tan{\varphi}\right) \nonumber \\
& & + \frac{2}{3}\ E_{n,m}^{\rm shift}\left(\psi,3A,0,\psi\tan{\varphi}\right).
\end{eqnarray}

We consider also a 4-chain structure (not pictured) with two non-equivalent 
sets of particles.
For symmetry reason the distances between rows 1-2 and 3-4 must be the same
and equal to $(\psi/2) \tan\varphi_1$ while the distance between rows 2-3
$(\psi/2) \tan\varphi_2$ may be different.
The optimization of the energy for $A=4$ implies the angle values
$\varphi_1=0.33385 \pi$ and $\varphi_2=0.33328 \pi$ which are very close
to one another. 
To simplify numerical calculations, we set $\varphi_1=\varphi_2=\varphi$;
the change in energies appears at the fourth decimal digit.
Respecting that
\begin{equation} \label{aaa}
4 A = \psi a^2 ,
\end{equation} 
the energy of the 4-chain structure reads as
\begin{eqnarray} 
E^{\rm 4ch}_{n,m}(\varphi,\psi,A) & = & E_{n,m}^{\rm rec}(\psi,4A)
+ E_{n,m}^{\rm shift}\left(\psi,4A,0,\psi\tan{\varphi}\right)
\nonumber \\ & & + \frac{3}{2} 
E_{n,m}^{\rm shift}\left(\psi,4A,\frac{1}{2},\frac{\psi}{2}\tan{\varphi}\right)
\nonumber \\ & & + \frac{1}{2} 
E_{n,m}^{\rm shift}\left(\psi,4A,\frac{1}{2},\frac{3\psi}{2}\tan{\varphi}\right). 
\label{e4}
\end{eqnarray}

One can proceed to simplified equidistant more-chain structures by considering
only two independent variables $\psi$ and $\varphi$, but increasing the number 
of chains the derivation of the energy per particle becomes more cumbersome. 

\section{Standard $12-6$ LJ potential} \label{Sec3}
In this section, we study the standard $12-6$ case of the LJ 
potential (\ref{n-m}). 
\subsection{Simple Bravais lattices}
\begin{figure}[htb]
\begin{center}
\includegraphics[clip,width=0.7\textwidth]{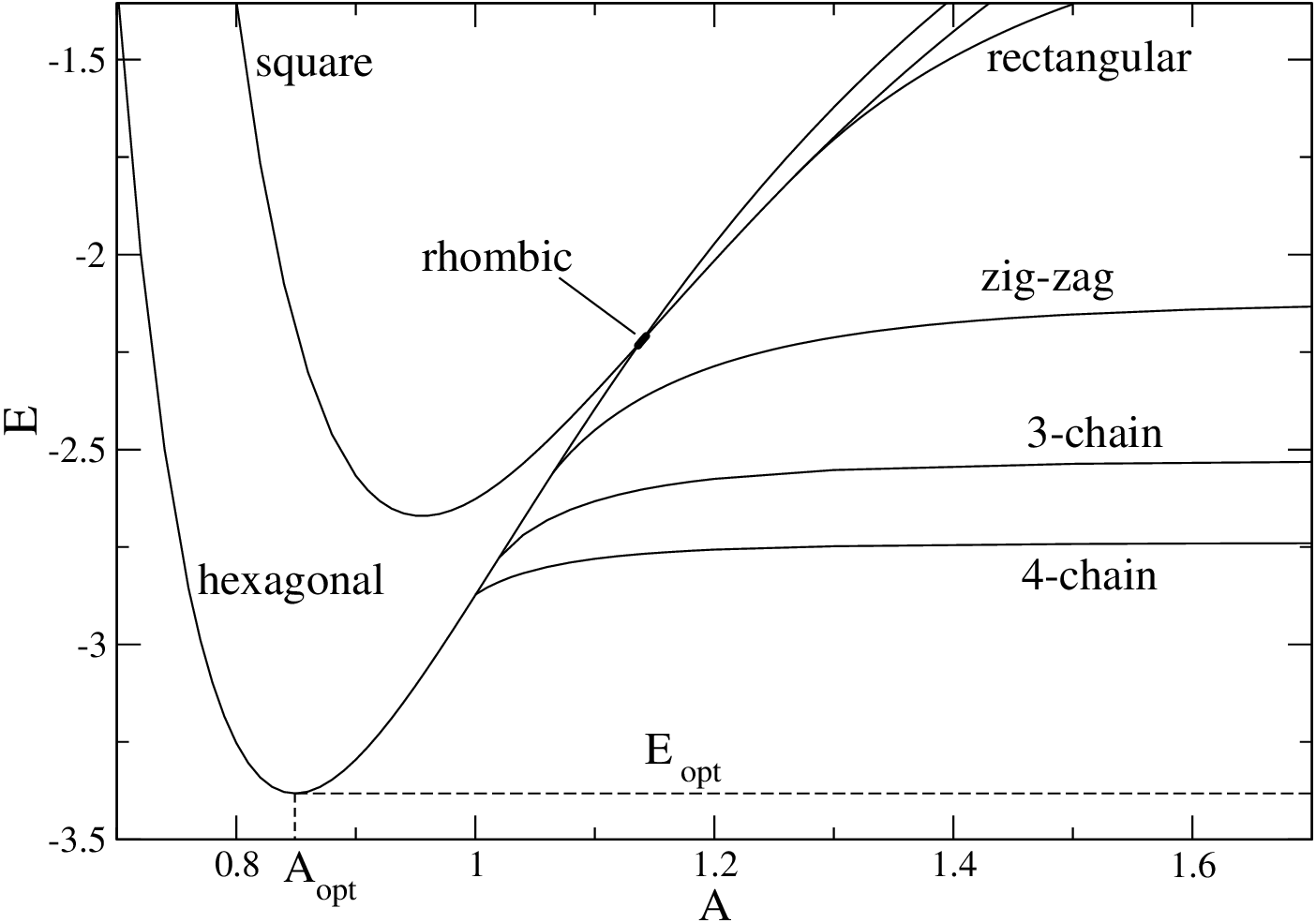}
\caption{$12-6$ LJ potential: 
The energy per particle as a function of $A$ for simple Bravais lattices, 
zig-zag (2-chain), 3-chain and 4-chain structures.
The optimal hexagonal energy is indicated by the horizontal dashed line.}
\label{fig:ea}
\end{center}
\end{figure}

\begin{figure}[htb]
\begin{center}
\includegraphics[clip,width=0.7\textwidth]{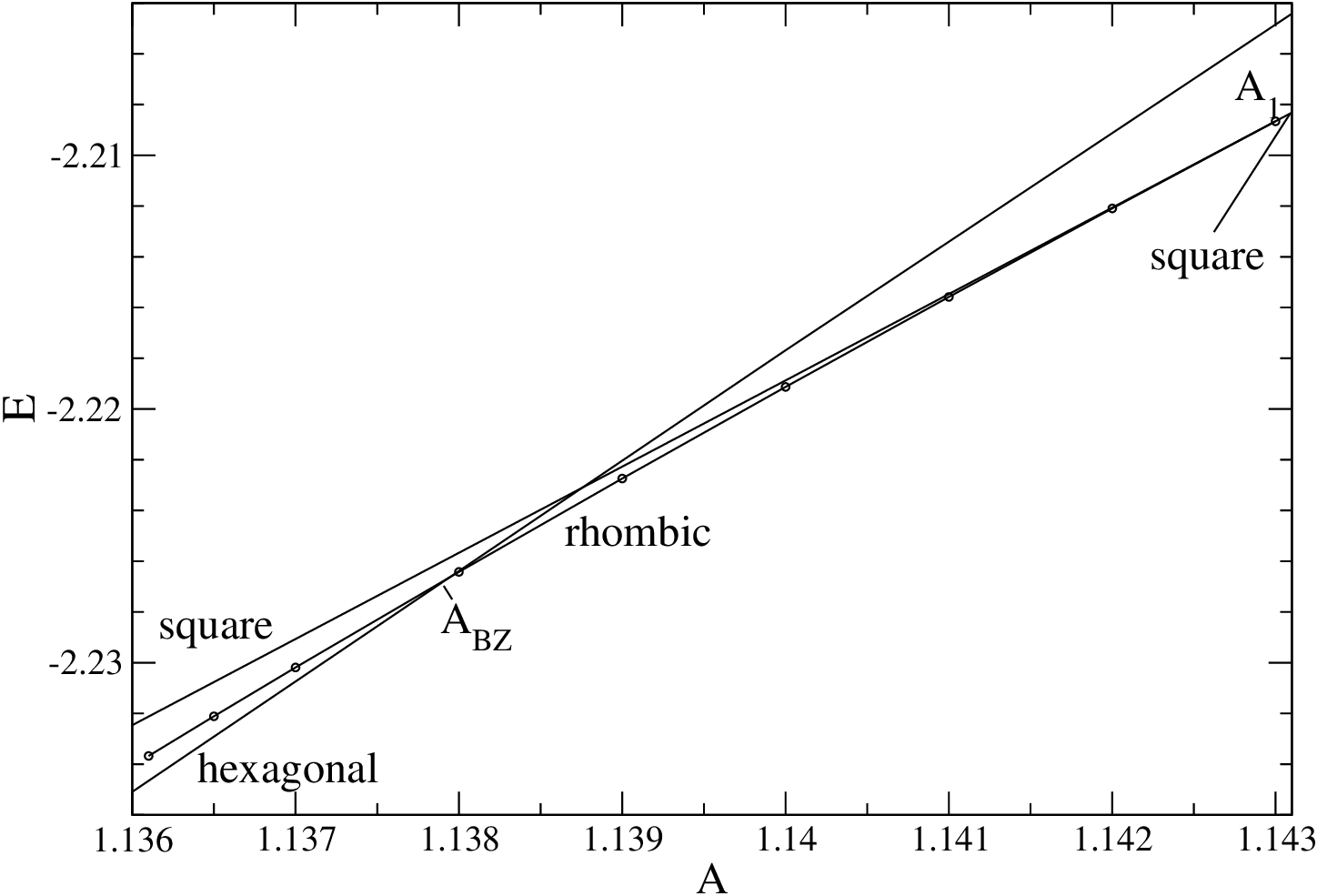}
\caption{Zoom of a short $A$-interval from the previous figure where 
various Bravais structures determine the ground state. 
The rhombic structure (small circles) becomes dominant in the interval 
$(A_{\rm BZ}, A_1)$.}
\label{fig:ead}
\end{center}
\end{figure}

Restricting himself to simple Bravais lattices, B\'etermin \cite{Bet} 
showed that by increasing the value of $A$, the hexagonal, 
rhombic, square and rectangular lattices become successively energy minimizers.
The transition between the hexagonal and rhombic lattices was found at 
$A_{\rm BZ}\approx 1.1378475$, the transition between the rhombic and square 
lattices at $A_1\approx 1.1430032$ and finally the transition between 
the square and rectangular lattices at $A_2\approx 1.2679987$.
Our high-precision calculations give $A_{\rm BZ}\approx 1.13784740849$, 
$A_1\approx 1.14300316307$ and $A_2\approx 1.26799868096$, meaning a minor 
change in the last digit of  $A_{\rm BZ}$ and fully confirming the rest.
The transitions at $A_1$ and $A_2$ are of second order, i.e. the corresponding
parameters $\delta$ and $\Delta$ change continuously from the value 1 at
the critical point. 
The change of $\delta$ at $A_{\rm BZ}$ is discontinuous, which means 
the first-order transition between the hexagonal and rhombic structures.
The $A$-dependence of the ground-state energies for simple Bravais lattices 
is pictured in Fig. \ref{fig:ea}, a zoom around the short interval 
$(A_{\rm BZ}, A_1)$ where the rhombic structure becomes energy minimizer 
is given in Fig. \ref{fig:ead}.

As is seen in Fig. \ref{fig:ea}, the energy of the hexagonal lattice
exhibits a minimum at $A_{\rm opt}$ which corresponds to the previously
defined optimal hexagonal lattice.
Recalling that the hexagonal lattice is a special case of the rhombic
one, namely $E^{\rm hex}_{12,6}(A) = E_{12,6}^{\rm rh}(\delta=\sqrt{3},A)$,
and using the representation (\ref{enmrh}), one has
\begin{equation}
E^{\rm hex}_{12,6}(A) = \frac{v_{12}^{\rm rh}(\sqrt{3})}{A^6}
- 2 \frac{v_6^{\rm rh}(\sqrt{3})}{A^3} .
\end{equation}
The energy minimization with respect to $A$ implies 
\begin{equation} \label{aopt}
A_{\rm opt} = 
\left[\frac{v_{12}^{\rm rh}(\sqrt{3})}{v_6^{\rm rh}(\sqrt{3})}\right]^{1/3}
\approx 0.8491235647 .
\end{equation}
Using the hexagonal relation $A=a^2\sqrt{3}/2$, the optimal lattice
spacing $a_{\rm opt}\approx 0.990193636287$ is very close to 1 which
is the distance between two isolated particles corresponding to 
the energy minimum. 
The optimal energy per particle
\begin{equation} \label{eopt}
E_{\rm opt} = E_{12,6}^{\rm rh}(\sqrt{3},A_{\rm opt}) \approx -3.38212347861
\end{equation}
is plotted in Fig. \ref{fig:ea} as a horizontal dashed line. 
It applies for $A>A_{\rm opt}$ where only a fraction of the available space 
is occupied by the optimal hexagonal lattice.
For $A\le A_{\rm opt}$, the space is fully occupied by the hexagonal 
lattice whose energy per particle depends on $A$ (see the solid curve 
in Fig. \ref{fig:ea}).

Returning to periodic Bravais lattices, for very large $A$ the rectangles 
become extremely narrow \cite{Bet} which means that also the aspect 
ratio $\Delta\to\infty$, see the illustration in the left panel 
of Fig. \ref{f2}. 
It can be readily shown that in that limit the most relevant terms from 
the Misra series representation of ${v_n^{\rm rec}(\Delta)}$ in equation 
(\ref{vnDelta}) correspond to the sum
$\sum_{j=1}^{\infty} z_{n/2+1}(j^2/\Delta)/(\frac{n}{2}-1)!$.
Consequently,
\begin{eqnarray}
E_{12,6}^{\rm rec}(\Delta,A) & = &  \frac{v_{12}^{\rm rec}(\Delta)}{A^6} -
2 \frac{v_6^{\rm rec}(\Delta)}{A^3} \nonumber \\ & \mathop{\sim}_{A\to\infty} & 
\frac{1}{120 A^6} \sum_{j=1}^{\infty} z_7\left( \frac{j^2}{\Delta}\right)
- \frac{1}{A^3} \sum_{j=1}^{\infty} z_4\left( \frac{j^2}{\Delta}\right) .
\end{eqnarray}
Since
\begin{equation}
z_4(y) \mathop{\sim}_{y\to 0} \frac{2}{y^3} + O(1) , \qquad
z_7(y) \mathop{\sim}_{y\to 0} \frac{120}{y^6} + O(1) ,
\end{equation}
we end up with
\begin{equation} \label{energy}
E_{12,6}^{\rm rec}(\Delta,A) \mathop{\sim}_{A\to\infty}
\left( \frac{\Delta}{A} \right)^6 \zeta(12) -
2 \left( \frac{\Delta}{A} \right)^3 \zeta(6) , 
\end{equation}
where $\zeta(s) = \sum_{j=1}^{\infty} 1/j^s$ is the Riemann's zeta function.
Recalling that $\Delta/A = 1/a^2$ the energy per particle (\ref{energy}) 
corresponds to the one-dimensional array of particles with the $12-6$ 
LJ potential; the point is that for $\Delta\to\infty$ all other
particles are at infinite distance from the reference one.
For fixed $A$, the energy (\ref{energy}) as a function of $\Delta$
is minimal for
\begin{equation} 
\Delta^* \mathop{\sim}_{A\to\infty} \left[ \frac{\zeta(6)}{\zeta(12)}\right]^{1/3}
A = 1.00566397 A .
\end{equation}
The corresponding energy per particle is finite:
\begin{equation} \label{energyA}
E_{12,6}^{\rm rec}(\Delta^*,A) \mathop{\sim}_{A\to\infty} 
- \frac{\zeta(6)^2}{\zeta(12)} = -1.034732272 .
\end{equation}

\subsection{Second-order transitions between rhombic, square and rectangular 
phases} 
First we study the second-order transition between the rhombic and square 
phases.
Parameterizing $\delta=\exp({\epsilon})$, the symmetry $\delta\to 1/\delta$ 
of the energy (\ref{enmrh}) is equivalent to the transformation 
$\epsilon\to -\epsilon$ and the energy is an even function of $\epsilon$.
The Ginsburg-Landau form of its expansion around $\epsilon=0$ reads as
\begin{equation} \label{veps2}
E_{12,6}^{\rm rh}(\epsilon,A) = E_{12,6}^{\rm rh}(0,A) + g_2(A)\epsilon^2 
+ g_4(A)\epsilon^4 + \ldots ,
\end{equation}
where the Misra representation of the expansion functions $g_2(A)$ and 
$g_4(A)$ is at our disposal (we do not write them explicitly because
formulas are too lengthy). 
The critical point $A_1$ is identified with the condition $g_2(A_1)=0$.
The functions $g_2(A)$ and $g_4(A)$ behave in the vicinity of the critical 
point $A_1$ as follows
\begin{eqnarray}
g_2(A) & = & g_{21}(A_1-A) + {\cal O}[(A_1-A)^2], \nonumber \\
g_4(A) & = & g_{40}
+ {\cal O}(A_1-A), \label{g2g4}
\end{eqnarray}
where the constant prefactors $g_{21}<0$ and $g_{40}>0$. 
The minimum energy is reached at $\epsilon^*\approx\delta^*-1$ given by
\begin{equation} \label{epsceq}
\frac{\partial}{\partial \epsilon} E_{12,6}^{\rm rh}(\epsilon,A)
\big\vert_{\epsilon=\epsilon^*}
\approx 2 g_2(A)\epsilon^* +4 g_4(A)(\epsilon^*)^3 = 0 .
\end{equation}
For $A>A_1$, there is only one trivial solution $\epsilon^*=0$ which
corresponds to the square lattice.
For $A<A_1$, besides the trivial solution $\epsilon^* = 0$ one gets
also two non-trivial conjugate solutions $\pm \epsilon^*$ with
\begin{equation} \label{epsc}
\epsilon^* = \left(-\frac{g_2(A)}{2 g_4(A)}\right)^{1/2}
\approx \left(-\frac{g_{21}}{2 g_{40}}\right)^{1/2} \sqrt{A_1-A} , 
\end{equation}
which provide a lower energy than the trivial solution.
The order parameter $\epsilon^*\propto\sqrt{A_1-A}$ is thus associated
with the mean-field critical index $\beta = 1/2$. 
The fitting of the numerical dependence of $\delta^*-1$ on $A_1-A$ 
gives $\beta=0.5003$, confirming the mean-field character of 
the phase transition.

Analogously one can find the same type of transition between square and 
rectangular phases, using the energy expression (\ref{enmrec}) and 
the parameterization $\Delta=\exp({\epsilon'})$ ensuring the symmetry
$\epsilon'\to -\epsilon'$ of the energy.
The numerical value $\beta = 0.5006$ confirms the same scenario and 
the mean-field character of this transition for $A\to A_2^+$.

\subsection{Beyond simple Bravais lattices}
Including the double-periodic zig-zag (2-chain) structure
modifies the ground-state phase diagram significantly. 
As one can see in Fig. \ref{fig:ea}, the hexagonal lattice remains 
the energy minimizer for low values of $A$, namely $0<A<A_t=1.0650263$.
But for $A>A_t$ that is the zig-zag structure which minimizes the energy 
and all other Bravais lattices become suppressed.
The phase transition between the hexagonal and zig-zag structures is of 
first order, as the parameters $\varphi$ and $\psi$ undergo a stepwise change 
at $A=A_t$, namely from the hexagonal values $\varphi=\pi/3=1.0471975511966$ 
and $\psi=1/\sqrt{3}=0.57735026918963$ to the zig-zag values
$\varphi=1.03912932433$ and $\psi=0.5313649234$.

Let us study the limiting case $A\to\infty$ of the zig-zag phase.
Numerical calculations up to $A=5$ reveal that 
$A \psi\to {\rm const}\approx 0.429294$, i.e. $\psi\to 0$ when $A\to\infty$.
This corresponds to elongated zig-zag strips as counterparts of
the elongated rectangles, see Fig. \ref{f2}.
In analogy with the elongated rectangles it can be shown that in the limit
$\psi\to 0$ the leading terms of the zig-zag energy (\ref{ezigzag}) correspond 
to the triangle row adjacent to the reference site:
\begin{eqnarray} \label{ezigzaglim}
E^{\rm zz}_{12,6}(\varphi,\psi,A)  &\mathop{\sim}_{A\to\infty}&
\frac{\zeta(12)}{(2A\psi)^6}-\frac{2\zeta(6)}{(2A\psi)^3}
-\frac{1}{(2A\psi)^3}\sum_{j=-\infty}^{\infty}
\frac{1}{(j^2+j+\frac{1}{4\cos^2{\varphi}})^3} \nonumber\\
& & + \frac{1}{2(2A\psi)^6}\sum_{j=-\infty}^{\infty}
\frac{1}{(j^2+j+\frac{1}{4\cos^2{\varphi}})^6} .
\end{eqnarray}
The first two terms on the r.h.s. sum over the line on which the reference
site belongs, in analogy with the rectangular equation (\ref{energy}),
with the substitution $A\to2A$ and $\Delta\to 1/\psi$.
The last two terms sum over the upper line of the triangle array.
They are given by the $k=0$ case of last sum in equation (\ref{s2}) with 
$c_1=1/2$ and $c_2=(\psi\tan{\varphi})/2$.
The sums in equation (\ref{ezigzaglim}) are slowly converging, but 
they can be summed up analytically by using a generating sum, 
see formulas (\ref{gen})-(\ref{sum3}) in Appendix.
Minimizing the energy (\ref{ezigzaglim}) with respect to $\varphi$ and $\psi$
yields 
\begin{eqnarray} 
\varphi^* & \mathop{\sim}_{A\to\infty} & 1.0500805037 = 0.33425100567 \pi ,
\nonumber \\
\psi^* & \mathop{\sim}_{A\to\infty} & \frac{0.42929422368}{A}. \label{solve}
\end{eqnarray}
Note that the angle $\varphi^*$ is close to $\pi/3$ for the present $12-6$ 
potential which means that the rows are composed of almost equilateral 
triangles.
Within the calculations presented in the next section, for the $n-6$ potential 
with very large $n=42$ we get $\varphi^*=0.33349 \pi$. 
Therefore, it is likely in the limit $n\to\infty$ that $\varphi^*=\pi/3$ 
holds exactly. 
In other words, in the low-density limit $A\to\infty$ the system seems 
to produce the zig-zag structure with distant rows of optimal equilateral 
triangles.
The corresponding energy per particle
\begin{equation} \label{energyzz}
E_{12,6}^{\rm zz}(\varphi^*,\psi^*,A) \mathop{\sim}_{A\to\infty} -2.1163
\end{equation}
is well below the one for elongated rectangles (\ref{energyA}).

As is clear from Fig. \ref{fig:ea}, the 3-chain and 4-chain structures 
become dominant at intermediate and large values of $A$ and provide
even lower energies than the zig-zag phase.
The phase transitions to the hexagonal phase are of first order.
Performing the large-$A$ analysis for 3-chains and 4-chains, 
the corresponding energies 
\begin{equation} \label{energychain}
E_{12,6}^{\rm 3ch}(A) \mathop{\sim}_{A\to\infty} -2.5279 , \qquad
E_{12,6}^{\rm 4ch}(A) \mathop{\sim}_{A\to\infty} -2.7392
\end{equation}
are lower than the one for the zig-zag structure, see also Fig. \ref{fig:ea}.
By increasing the number $n$ of chains to infinity one might expect that
the energy goes to the optimal hexagonal one.
Applying the $[1/1]$ Pad\'e extrapolation
\begin{equation} \label{Pade}
E_{12,6}^{\rm n-chain}(A\to\infty) = \frac{a_0+a_1 n}{b_0+n}
\end{equation}
to $n=2$ (\ref{energyzz}) and $n=3,4$ (\ref{energychain}) data,
one obtains the value
\begin{equation} \label{infinity}
E_{12,6}^{\infty} = a_1 = - 3.3961 
\end{equation}
which is indeed very close to the optimal hexagonal energy (\ref{eopt}).

\section{General $n-6$ case} \label{Sec4}
\begin{figure}[htb]
\begin{center}
\includegraphics[clip,width=0.7\textwidth]{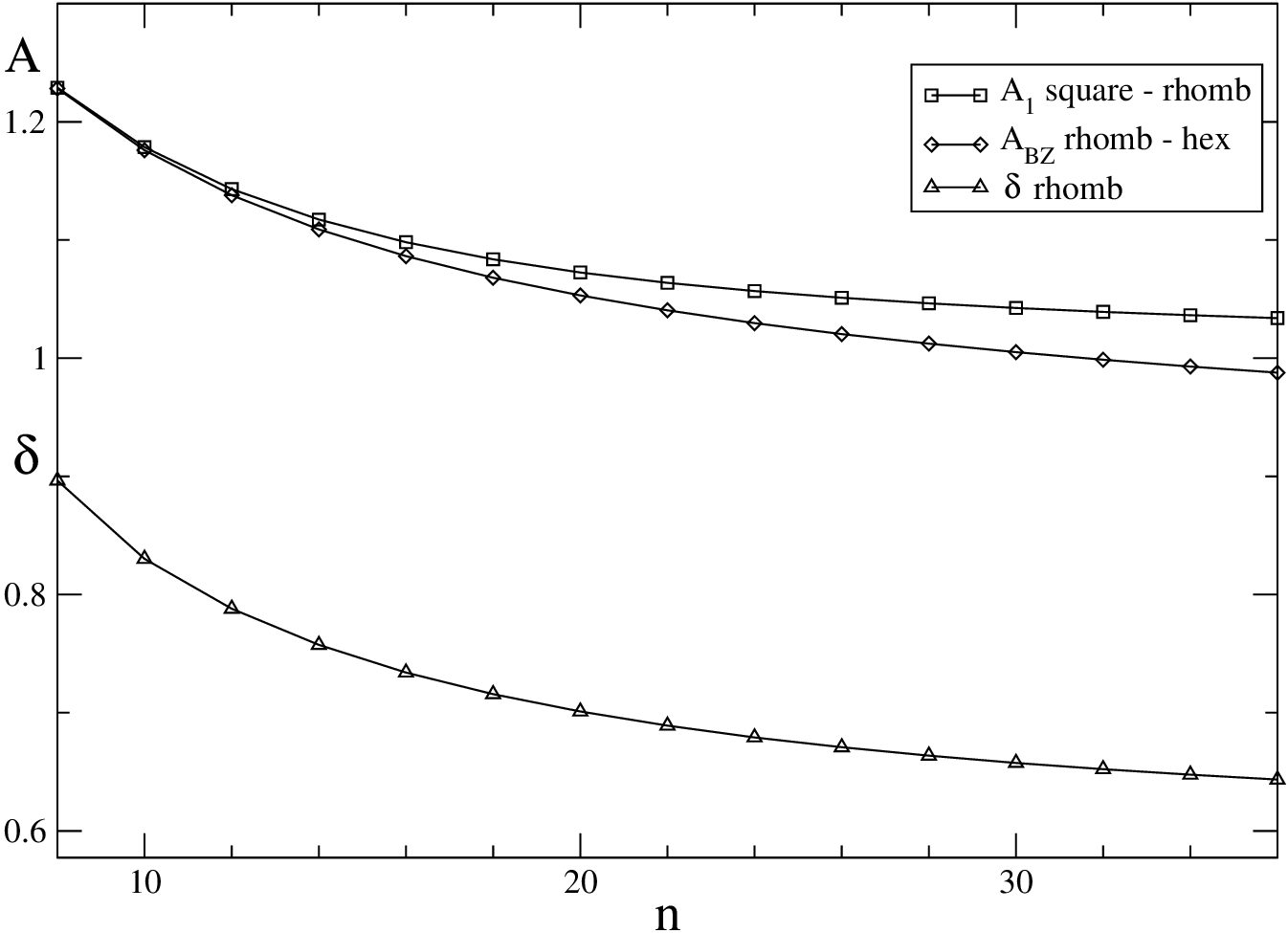}
\caption{The values of $A$ and $\delta$ at structural phase transitions 
between square-rhombic ($A_1$, squares) and rhombic-hexagonal 
($A_{\rm BZ}$, diamonds) structures as a function of even integer $n$.} 
\label{fig:adn}
\end{center}
\end{figure}

For the considered Bravais lattices, we analyze the $n-6$ interaction 
with even $n$ ranging from 8 to 36. 
The results are presented graphically in Fig. \ref{fig:adn}. 
As concerns the transition between the hexagonal and rhombic structures
at $A_{\rm BZ}$ (diamonds), there was a possibility that starting from some $n$ 
the first-order phase transition becomes of second order, but this is
not the case.
In the hard-core $n\to\infty$ limit, $A_{\rm BZ}(n)$ approaches to 
the close-packed hexagonal value $A_h=\sqrt{3}/2\approx 0.8660254$ and
the rhombic parameter $\delta$ goes to its hexagonal value 
$1/\sqrt{3}\approx 0.57735$ which was chosen as the bottom value of 
the $y$-axis in Fig. \ref{fig:adn}. 
The rhombic-square transition value $A_1(n)$ (squares) reaches asymptotically 
the closed-packed square value 1.

\begin{figure}[htb]
\begin{center}
\includegraphics[clip,width=0.7\textwidth]{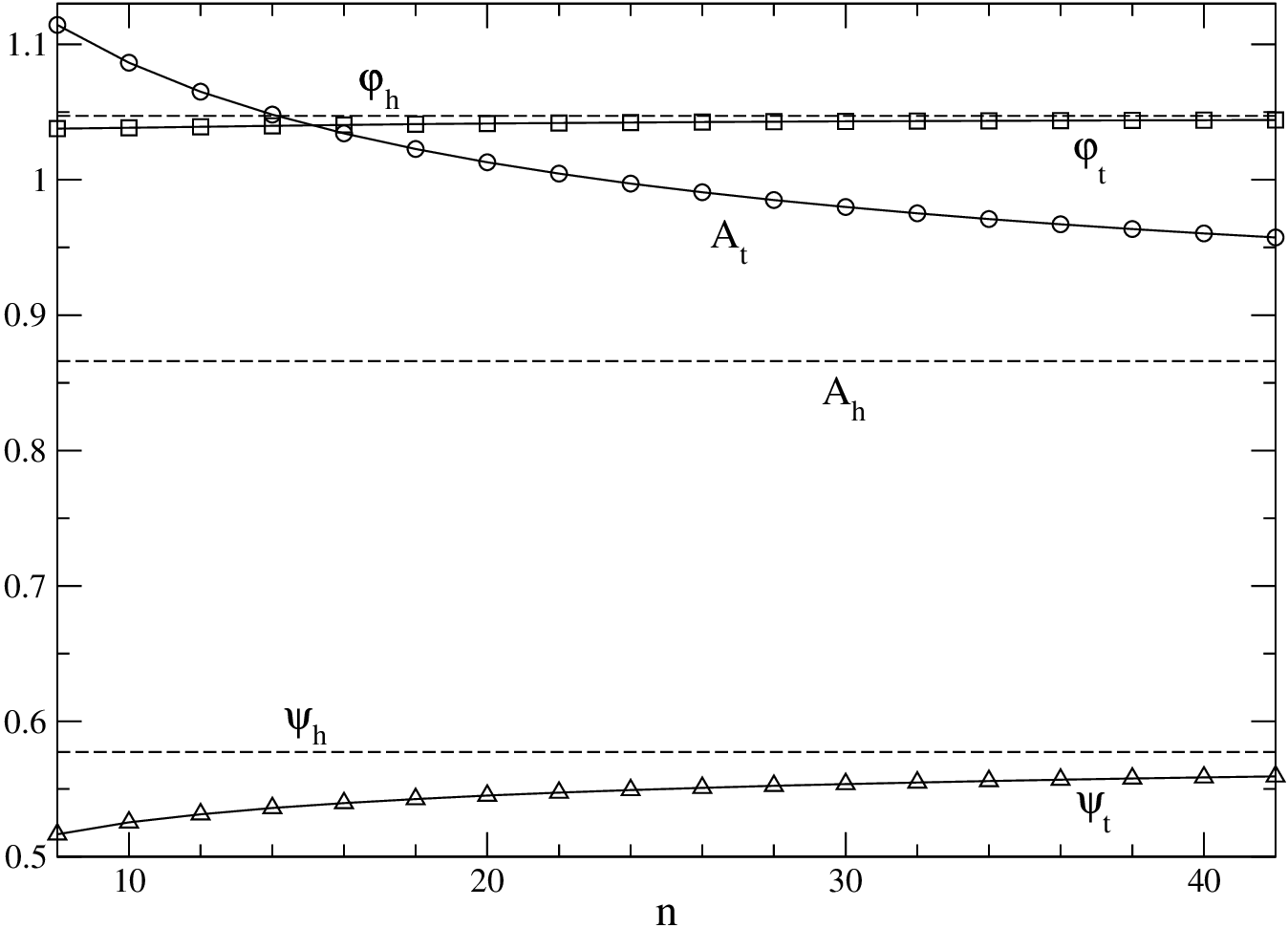}
\caption{The first-order phase transition between hexagonal and zig-zag 
structures for even $n$ from 8 to 42. 
The transition values of $A_t$ (circles), $\varphi_t$ (squares) and 
$\psi_t$ (triangles) reach in the $n\to\infty$ limit their close-packed 
hexagonal values $A_h$, $\varphi_h$ and $\psi_h$, respectively, 
denoted by the horizontal dashed lines.}
\label{fig:efp}
\end{center}
\end{figure}

For any finite $n$, there is a first-order phase transition at $A_t(n)$
between the hexagonal lattice, which is energy minimizer for $A<A_t(n)$, 
and the zig-zag structure, which is energy minimizer for $A>A_t(n)$, 
see circles in Fig. \ref{fig:efp}. 
In the limit $n\to\infty$, the close-packed hard disks form a hexagonal 
lattice with $A = A_h$ (horizontal dashed line). 
There is no phase for $A<A_h$ due to the hard-core restrictions among
particles.
The hexagonal phase, which exists only at $A_h$, converts immediately to 
the zig-zag phase which is the energy minimizer for all $A>A_h$.
The corresponding transition values of the lattice parameters $\varphi_t$
(squares) and $\psi_t$ (triangles) go to their hexagonal values 
$\varphi_h=\pi/3$ and $\psi_h=1/\sqrt{3}$.

Now we analyze in detail the limit $n\to\infty$ for the zig-zag structure
to describe its transformation to the hexagonal lattice of hard disks 
with diameter 1 at $A_h$. 
In that limit, one keeps in the energy (\ref{ezigzag}) only the attractive 
part of the interaction, whereas the repulsive part of interaction 
is replaced by a set of hard-core constraints.
These constraints can be established by simple geometrical considerations, 
namely the three nearest neighbors of a chosen particle cannot be at distance 
smaller than 1 from that particle.
This implies the conditions
\begin{eqnarray}
\psi &\ge& \frac{1}{2A} , \nonumber \\
\frac{\psi}{\cos{\varphi}^2} &\ge & \frac{2}{A} , \nonumber \\
\frac{\psi^2}{4}+\left(1-\frac{\psi}{2}\tan{\varphi}\right)^2
&\ge& \frac{\psi}{2A}. \label{cons}
\end{eqnarray}
With these constraints, the minimization of the attractive part of 
the zig-zag energy for values of $A$ slightly above $A_h$ showed
that only the first constraint remains to be a strict inequality $\psi>1/2A$,
while the other two become the equalities.
Thus we can write down the set of two equations for three free parameters
$A$, $\varphi$ and $\psi$:
\begin{eqnarray}
\frac{\psi}{\cos^2\varphi} & = & \frac{2}{A} , \nonumber \\
\frac{\psi^2}{4}+\left( 1-\frac{\psi}{2} \tan{\varphi}\right)^2
& = & \frac{\psi}{2A}.
\end{eqnarray}
Taking $\varphi$ as an independent parameter, one gets
\begin{equation} \label{cons2}
A = \sin{2\varphi} , \qquad \psi = \cot{\varphi}.
\end{equation}
For $\varphi=\varphi_h=\pi/3$, one can check that we get the close-packed
hexagonal values $A_h=\sqrt{3}/2$ and $\psi_h=1/\sqrt{3}$.
Based on (\ref{cons2}), for $A$ close to $A_h$ we get the expansions for 
small deviations of $\varphi$ and $\psi$ from their hexagonal
values in powers of $A-A_h$,
\begin{eqnarray}
\varphi-\varphi_h &=& A-A_h+\sqrt{3}(A-A_h)^2-\frac{2}{3}(A-A_h)^3
+{\cal O}(A-A_h)^4 , \nonumber \\
\psi-\psi_h&=&\frac{4}{3}(A-A_h)+\frac{16}{3\sqrt{3}}(A-A_h)^2
+\frac{112}{9}(A-A_h)^3+{\cal O}(A-A_h)^4 . \nonumber\\ \label{cons3}
\end{eqnarray}
These expansions were checked also numerically.
The fact that the expansions are analytic in the deviation $A-A_h$
implies that there is no phase transition in the limit of close-packed 
hard disks.
We conclude that the first-order transition between the hexagonal and 
zig-zag phases, which exists for any finite $n$, disappears for $n\to\infty$. 

A similar analysis can be made for 4-chain structures with qualitatively 
similar results.

\section{Conclusion} \label{Sec5}
In this paper, we have studied the 2D ground-state Wigner structures of
classical particles with standard $12-6$ and $n-6$ LJ 
pairwise interactions.
We have used special lattice summation techniques \cite{ST,TS2,TS}
which permit one to express the energy per particle as a quickly converging
series of Misra functions.
The main aim was to extend B\'etermin's treatment of simple Bravais
lattices \cite{Bet} to periodic 2D lattices with more than one particle 
per elementary cell.

Fixing the area per particle $A$, we have introduced in section \ref{Sec2}
four simple Bravais lattices (see Fig. \ref{f1}) and the extensions of
the rectangular (1-chain) lattice to double-periodic zig-zag (2-chain),
3-chain and 4-chain structures (Fig. \ref{f2}).

Section \ref{Sec3} deals with the standard $12-6$ LJ potential. 
We have confirmed the results of B\'etermin \cite{Bet} that as $A$ increases
the hexagonal, rhombic, square and rectangular phases become successively
the energy minimizers.
The second-order transitions between the rhombic/square and square/rectangular 
phases are shown to be of mean-field type.
As is seen in Fig. \ref{fig:ea}, the inclusion of the zig-zag (2-chain), 
3-chain and 4-chain structures modifies the phase diagram substantially. 
These phases become dominant at intermediate and large values of $A$
and their phase transitions to the hexagonal phase are of first order.
We anticipate that as the number of chains goes to infinity, this
transition will be of second order.
The limit $A\to\infty$ can be treated analytically and we found that the
energy of the zig-zag phase (\ref{energyzz}) is well below the rectangular 
one (\ref{energyA}); the rectangular phase is the energy minimizer 
of simple Bravais lattices for $A\to\infty$. 
The corresponding results for the 3-chain and 4-chain structures
(\ref{energychain}) indicate a systematic decrease of the energy
to the one of the phase separated state (the optimal hexagonal lattice)
as the number of chains increases; the Pad\'e extrapolation (\ref{Pade}) 
of available $n=2,3,4$ data implies the result (\ref{infinity}) which 
is indeed close to the optimal value (\ref{eopt}).

The general $n-6$ ($n$ being an even integer) LJ potential is 
the subject of section \ref{Sec4}; here, we restrict ourselves to
Bravais and zig-zag structures.
In the case of simple Bravais lattices $n$ ranges from 8 to 36 
(Fig. \ref{fig:adn}), while the first-order transition between 
the hexagonal and zig-zag phases is studied for $n$ ranging from 8 to 42 
(Fig. \ref{fig:efp}).
We studied analytically the interesting limit $n\to\infty$, where
the hexagonal closed-packed lattice of hard disks exists only at 
$A_h=\sqrt{3}/2$ (there is no phase for $A<A_h$ due to the hard-core 
restrictions among the particles) and the zig-zag phase dominates in 
the whole region $A>A_h$.
It was shown that the parameters of the zig-zag phase are analytical functions
of the deviation $A-A_h$, so its transformation to the hexagonal phase at 
$A_h$ is smooth and there is no phase transition for hard disks.
In other words, the first-order phase transition between the hexagonal
and zig-zag structures, which exists for any finite value of $n$, disappears
for $n\to\infty$. 

\ack
The support received from VEGA Grant No. 2/0003/18 is acknowledged. 

\renewcommand{\theequation}{A.\arabic{equation}} 
\setcounter{equation}{0}

\section*{Appendix}
The first step of the method developed in \cite{ST,TS2,TS} is 
the replacement of the interaction term of the type $1/r^n$ by using
the gamma identity:
\begin{equation} \label{transf}
\frac{1}{r^n}=\frac{1}{(n-1)!} \int_{0}^{\infty} {\rm d}t\ 
t^{n-1} \exp(-r t) .
\end{equation}
A series of subsequent transformations leads to the representation
of the energy per particle with the aid of Misra functions \cite{Misra}:
\begin{equation} \label{Misra}
z_{\nu}(y) = \int_0^{1/\pi} \frac{{\rm d}t}{t^{\nu}} 
\exp\left( -\frac{y}{t}\right) , \qquad y>0 . 
\end{equation}
For even powers $n$ or $m$ of $r=\sqrt{r_x^2+r_y^2}$ one gets the Misra
functions with an integer values of $\nu$; their representation in terms
of special functions, suitable especially in MATHEMATICA, is easy to obtain. 
We cut the Misra series representation at the eighth term to get the precision 
of the energy more than 25 digits.
In particular, for the case $12-6$ one needs the following Misra functions
\begin{eqnarray} \label{znuasymp}
z_{-1}(y) & = & \frac{1}{2}\left[\frac{e^{-\pi{y}}(1-\pi y)}{\pi^2}+y^2
\Gamma(0,\pi y)\right], \nonumber\\
z_{-4}(y) & = & \frac{{\rm e}^{-\pi{y}}}{120\pi^5}
\left(24-6\pi y+2\pi^2 y^2-\pi^3 y^3+\pi^4 y^4\right) \nonumber\\
& & - \frac{y^5}{120}\Gamma(0,\pi y), \nonumber\\
z_4(y) & = & \frac{{\rm e}^{-\pi{y}}\left(2+2\pi y+\pi^2 y^2\right)}{y^3},
\nonumber\\
z_7(y) & = & \frac{{\rm e}^{-\pi{y}}}{y^6}(120+120\pi y+60\pi^2 y^2+20\pi^3 y^3
\nonumber\\ & & +5\pi^4 y^4+\pi^5 y^5) ,
\end{eqnarray}
where $\Gamma(x,y)$ stands for the incomplete Gamma function \cite{Gradshteyn}.

The function $v_n^{\rm rh}(\delta)$ in the energy representation (\ref{enmrh})
for the rhombic lattice and $v_n^{\rm rec}(\Delta)$ in the energy representation
(\ref{enmrec}) for the rectangular lattice read (for even values of $n$)
\begin{eqnarray} \label{vndelta}
v_n^{\rm rh}(\delta) & = & \frac{1}{2^{n/2+1}(\frac{n}{2}-1)!} \Bigg\{ 
4\pi^{n-1} \sum_{j=1}^\infty \left[ z_{2-n/2}\left(\frac{4j^2}{\delta}\right)
+z_{2-n/2}\left(4j^2\delta\right) \right] \nonumber\\
& & + \frac{\left(\frac{n}{2}+1\right)\pi^{n/2}}{
\frac{n}{2}\left(\frac{n}{2}-1\right)} + 4\sum_{j,k=1}^\infty
z_{n/2+1}\left(\frac{j^2}{\delta}+k^2\delta\right) \nonumber \\ & &
+4\pi^{n-1} \sum_{j,k=1}^\infty\left[1+(-1)^j (-1)^k\right]
z_{2-n/2}\left(\frac{j^2}{\delta}+k^2\delta\right)\nonumber\\ & & 
+ 2 \sum_{j=1}^\infty\left[ z_{n/2+1}\left(\frac{j^2}{\delta}\right)
+z_{n/2+1}\left(j^2\delta\right) \right] 
\nonumber\\ & & + 4 \sum_{j,k=1}^\infty
z_{n/2+1}\left[\frac{(j-1/2)^2}{\delta}+(k-1/2)^2\delta\right] \Bigg\} ,
\end{eqnarray}
\begin{eqnarray} \label{vnDelta}
v_n^{\rm rec}(\Delta) & = & \frac{1}{2(\frac{n}{2}-1)!}\Bigg\{2\pi^{n-1} 
\sum_{j=1}^\infty \left[ z_{2-n/2}\left(\frac{j^2}{\Delta}\right)
+z_{2-n/2}\left(j^2\Delta\right) \right] \nonumber\\ & &
+\frac{\pi^{n/2}}{\frac{n}{2}\left(\frac{n}{2}-1\right)}
+ 4\pi^{n-1} \sum_{j,k=1}^\infty z_{2-n/2}\left(\frac{j^2}{\Delta}+k^2\Delta\right)
\nonumber\\ & &
+ 2\sum_{j=1}^\infty\left[ z_{n/2+1}\left(\frac{j^2}{\Delta}\right)
+z_{n/2+1}\left(j^2\Delta\right)\right] \nonumber\\ & &
+ 4\sum_{j,k=1}^\infty z_{n/2+1}\left(\frac{j^2}{\Delta}+k^2\Delta\right)\Bigg\}.
\end{eqnarray}

Let us consider an infinite 2D rectangular lattice of sides $\psi a$ and $a$,
and a reference point shifted with respect to a lattice point by the vector 
$(c_1\psi a,c_2 a)$.
Summing the potential (\ref{n-m}) over all points of the rectangular lattice
with respect to the reference point in close analogy with \cite{TS2}, 
we obtain the energy
\begin{equation} \label{v2}
E_{n,m}^{\rm shift}\left(\psi,2A,c_1,c_2\right) = \frac{1}{n-m} \left[
m \frac{\Sigma_{n/2}(\psi,c_1,c_2)}{(2A)^{n/2}}
- n\frac{\Sigma_{m/2}(\psi,c_1,c_2)}{(2A)^{m/2}} \right] ,
\end{equation}
where, considering that $n/2$ or $m/2$ is an integer $N$,   
\begin{eqnarray} \label{s2}
\Sigma_N(\psi,c_1,c_2) & = & \frac{1}{2(N-1)!} 
\Bigg\{ 2\pi^{2N-1}\sum_{j=1}^{\infty} 
\bigg[\cos(2\pi j c_1) z_{2-N}\left(\frac{j^2}{\psi}\right) \nonumber\\
& & + \cos(2\pi j c_2)z_{2-N}\left(j^2\psi\right)\bigg] + \frac{\pi^N}{N-1}
\nonumber\\ & & + 4\pi^{2N-1}\sum_{j,k=1}^\infty \cos(2\pi j c_1)\cos(2\pi k c_2)
z_{2-N}\left(\frac{j^2}{\psi}+k^2\psi\right) \nonumber\\
& & + \sum_{j,k=-\infty}^\infty 
z_{N+1}\left[(j+c_1)^2\psi+\frac{(k+c_2)^2}{\psi}\right] \Bigg\}.
\end{eqnarray}

To express analytically the sums in (\ref{ezigzaglim}), one starts with
the generating formula
\begin{equation} \label{gen}
\sum_{j=-\infty}^{\infty}\frac{1}{j^2+j+\kappa}
= 2\pi\frac{\tan\bigg({\frac{\pi}{2}\sqrt{1-4\kappa}}\bigg)}{\sqrt{1-4\kappa}}.
\end{equation}
Taking consecutive derivatives with respect to $\kappa$ one gets 
the required power of the denominator on the l.h.s.
A minor complication is that after inserting 
$\kappa=1/(4\cos{\varphi}^2)$ we get
\begin{equation} \label{a9}
\sqrt{1-4\kappa} = {\rm i} \tan{\varphi} ,
\end{equation}
but applying the equality $\tan({\rm i}x)={\rm i} \tanh x$ in the numerator 
of the r.h.s. of equation (\ref{gen}) we get real expressions as expected, 
e.g.
\begin{eqnarray}
\sum_{j=-\infty}^{\infty}\frac{1}{[j^2+j+1/(4\cos{\varphi}^2)]^3}
= -2\pi\cot^3{\varphi} \Bigg\{
-6\cot^2{\varphi}\tanh\left({\frac{\pi}{2}\tan{\varphi}}\right) \nonumber\\
\qquad +\pi\cosh^{-2} \left( \frac{\pi}{2}\tan{\varphi} \right)
\left[ 3\cot{\varphi}+\pi\tanh\left( {\frac{\pi}{2}\tan{\varphi}} \right)
\right] \Bigg\} . \label{sum3}
\end{eqnarray}

\end{document}